\documentstyle[prl,aps]{revtex}

\begin{document}
\draft

\title{Berry's Phase, Josephson's Equation, and the Dynamics of
       Weak Link Superconductors and Their Vortices}

\author{Frank Gaitan and Subodh R. Shenoy}
\address{International Center for Theoretical Physics; P. O. Box 586;
         Miramare; 34100 Trieste, ITALY}
\date{\today}

\maketitle

\begin{abstract}
We examine the dynamical consequences of Berry's phase for Josephson
junctions, junction arrays, and their vortices. Josephson's equation
and the related phase slip voltages are shown to be uneffected by
Berry's phase. In an annular Josephson junction, Berry's phase is seen to
generate a new current drive on a vortex.
In the continuum limit, vortex dynamics in a 2D array is shown to map
onto that of a 2D film.
A Hall sign anomaly is expected in arrays; and the merits of arrays
for studies of disorder on vortex motion is discussed.
\end{abstract}


\indent In a very interesting paper \cite{ath}, Ao and Thouless showed
that vortex motion in a Type-II superconductor generates a Berry
phase in the many-body ground state, and as a consequence, the Magnus
force of classical hydrodynamics acts on the vortex \cite{con}. Their
work led others to calculate the effective action for the
superconducting condensate $S_{\rm g}$ using the BCS model \cite{go3,vor}.
It was shown by one of these authors
\cite{vor} that when the Berry phase is non-removable, it manifests as
a Wess-Zumino (WZ) term in $S_{g}$. Ref.~\cite{vor} also pointed
out that a WZ term might arise in the effective action of a Josephson
Junction (JJ) when the superconducting gap phase $\phi$ contained a
moving vortex singularity.

In this Letter we examine the consequences of the Berry-phase-induced
WZ term for the dynamics of weak link superconductors and their
vortices. As far as we are aware, this is the first time such a study
has been carried out, either for single weak link systems, or for an array
of such links \cite{ram}. We work within the confines of the BCS model of
superconductivity; assume the gap phase contains a moving vortex
singularity $\phi({\bf r}-{\bf r}_{0}(t))$, where ${\bf r}_{0}(t)$ is
the vortex trajectory; and $T=0$. A detailed presentation of this work
will be given elsewhere \cite{us}. The essential results of this paper
are: (1) in a single superconducting grain, the Josephson-Gorkov equation
is not modified in any physically relevant way by the WZ term, and
consequently,
phase slip voltages are not effected by Berry's phase; (2) Berry's phase
leads to a new current drive acting on a vortex in an
annular JJ, though it proves ineffective (in this type of junction) for
geometrical reasons; (3) in a 2D Josephson Junction Array (JJA),
this new Berry-phase-induced current drive {\em is\/} operative; also,
in the Continuum Limit (CL), the superconducting (SC) dynamics of the
JJA maps onto that of a 2D film
so that a Hall Effect sign anomaly is expected in a JJA if the grains
are made of a material which shows such an anomaly in a 2D film; and
(4) a JJA in the CL appears ideally suited to test recent predictions of
Vinokur et.\ al.\ \cite{vin} regarding the Hall conductivity and
resistivity.

\underline{{\em (i) Single Superconducting Grain}:} We begin by considering
a single superconducting grain in which a
moving vortex singularity is present. The effective action for the
superconducting condensate $S_{g}$ was obtained in Ref.~\cite{go3}, though
our notation will follow Ref.~\cite{vor}. This action contains
space-time gradients up to second order and has the form $S_{g}=S_{0}+S_{wz}
+S_{2}$. $S_{0}$ corresponds to the action for a uniform gap function
and depends only on $|\Delta |$, whereas $S_{wz}$ and $S_{2}$ depend
{\em only\/} on the gap phase $\phi =\phi ({\bf r}-{\bf r}_{0}(t))$.
The gap equation corresponds to the Euler-Lagrange equation (EL-eqn)
obtained by variation of $S_{g}$ (viz.\ $S_{0}$) with respect to
$|\Delta |$ \cite{eck}. This equation acts to fix $|\Delta |$.
Outside the critical region near $T_{c}$, fluctuations away from the
extremal solution of the gap equation can be ignored, so one can
treat $|\Delta |$ as fixed, and approaching $\Delta_{0}$ far from the
vortex core. Beyond fixing $|\Delta |$, $S_{0}$ will not
interest us and so will not be written out explicitly. The remaining
contributions to $S_{g}$ are \cite{go3,vor}
\begin{equation}
S_{wz}+S_{2}=\int\, d^{2}xdt\,\left[ \,\rho_{s}\left(\frac{\hbar}{2}
              \partial_{t}\phi +eA_{0}\right) + N(0)\left(\frac{\hbar}{2}
               \partial_{t}\phi+eA_{0}\right)^{2} + \frac{m\rho_{s}}{2}
             {\bf v}_{s}^{2}+\frac{1}{8\pi}\left\{ \left(
              {\bf H}-{\bf H}_{ext}
              \right)^{2}-{\bf E}^{2}\right\} \right]\hspace{0.1in} .
                         \label{action}
\end{equation}
$S_{wz}$ is the WZ term and corresponds to the term in eqn.~(\ref{action})
whose integrand is gauge invariant and first order in time derivatives
of the gap phase $\phi$. $S_{2}$ is the remaining collection of terms. Here
$\rho_{s}$ is the density of superconducting electrons; $A_{0}$ is the
scalar potential induced by the vortex motion; $N(0)$ is the normal
density of states at the Fermi surface for one spin component; $m$ is the
electron mass; ${\bf v}_{s}=\hbar\left(\nabla\phi -(2e{\bf A}/\hbar c)
\right)/2m$ is the superfluid velocity; ${\bf H}=\nabla\times {\bf A}$ is the
microscopic magnetic field; ${\bf H}_{ext}=\nabla\times {\bf A}_{ext}$ is
the externally applied magnetic field; ${\bf E}
=-\nabla {\cal E} = -\nabla A_{0} - \partial ({\bf A}/c)/\partial t$ is
the electric field generated by the vortex motion; and ${\cal E}=-
\int^{{\bf x}}\, d{\bf l}\cdot {\bf E}$ is the gauge invariant
electromotive force whose gradient determines ${\bf E}$.

If we momentarily consider the case of a superconducting grain in which
no vortex is present, no WZ term will appear
in eqn.~(\ref{action}) \cite{vor}. This case was considered in
Ref.~\cite{eck} and the Josephson-Gorkov equation (JG-eqn) was obtained
by noting that the remaining term containing $\partial_{t}\phi$ is positive
definite and thus contributes least to $S_{g}$ when it is zero. It follows
from this remark that $-2eA_{0}=\hbar\partial_{t}\phi $ (viz.\ the
JG-eqn). Returning
to the case of a grain containing a moving vortex, here the WZ term is
present.
One can combine the integrands of the first two terms in eqn.~(\ref{action})
to obtain $\{ -\rho_{s}^{2}/4N(0)\} + N(0)\{ \hbar\partial_{t}(\phi /2)+
eA_{0}+
(\rho_{s}/2N(0))\}^{2}$. If one were to naively carry over the argument of
Ref.~\cite{eck}, one would argue that this expression was smallest
when the positive definite second term vanished, and subsequently, the
``JG-eqn'' (in the presence of a moving vortex) would be
$-2eA_{0}=\hbar\partial_{t}\phi ({\bf r}-{\bf r}_{0}(t)) +
   \rho_{s}({\bf r})/N(0)$.
On the basis of this analysis one would conclude that, for a given vortex
trajectory ${\bf r}_{0}(t)$, the scalar potential $A_{0}$
was modified as a result of the Berry-phase-induced WZ term. Because of
the intimate
connection between the JG-eqn and the voltage differences induced by vortex
motion through phase slip, such a modification, if true, might be thought
to lead to new physics. In fact, the above argument is not correct for the
case of a grain containing a moving vortex singularity, as we now show. The
essential point is that, when a moving vortex is present, ${\bf E}\neq 0$.
To obtain the JG-eqn, we vary $S_{g}$ with respect to ${\cal E}$
(recall ${\bf E}=-\nabla {\cal E}$) to obtain Poisson's equation
\begin{equation}
-\nabla^{2}{\cal E} = 4\pi \left[\, 2N(0)e\left\{\,\frac{\hbar}{2}
                       \partial_{t}\phi + eA_{0}+\frac{\rho_{s}}{2N(0)}
              \,\right\}\,\right] \hspace{0.1in} .
  \label{poisson}
\end{equation}
The term in square brackets on the RHS of this equation must equal the
electric
charge density $e(\rho_{s}-\rho_{0})$, where $\rho_{0}$ is the smeared out
charge density of the positive ions in the lattice. One
can obtain a consistency check on this identification by obtaining the
EL-eqn for $\phi$ which is the continuity equation for electric
charge. One finds \cite{us} that the quantity in square brackets in
eqn.~(\ref{poisson}) again appears as the charge density. Carrying out this
identification yields the JG-eqn (in the presence of a moving vortex),
$-2eA_{0}=\hbar\partial_{t}\phi +\rho_{0}/N(0)$.
Thus the JG-eqn does receive a correction term, although this correction is
a physically irrelevant constant, and will not effect voltage differences.
It can safely be discarded since the scalar potential is only defined
modulo a constant. Note that for the case considered by Ref.~\cite{eck},
no WZ term occurs, and $\nabla {\cal E}=0$; thus Poisson's equation will not
contain the final term in the curly brackets on the RHS of
eqn.~(\ref{poisson}), and the LHS will vanish. Thus, in this case, we
recover the usual JG-eqn from our approach. We also see that the argument
of Ref.~\cite{eck} (appropriate for the situation considered there) is
not generally applicable, being inconsistent with Poisson's equation when
a moving vortex is present.
Thus we obtain our first result: Berry's phase does {\em not\/}
modify the Josephson-Gorkov equation. Consequently, the usual arguments
concerning phase slip voltages are also unaltered \cite{and}. As first
pointed out in
Ref.~\cite{ath}, Berry's phase {\em does\/} effect the motion of a vortex
via the Magnus force. In Ref.~\cite{vor}, the derivation of this force from
the WZ term is given, and it appears that this is the only effect of
Berry's phase in the case of a single superconducting grain.

\underline{{\em (ii) Single JJ}:} We go on now to the case of a Josephson
Junction in
which two superconducting grains (L, R) are coupled through a weak link by
the Josephson effect. We assume the weak link to be a
tunneling barrier (TB). The Hamiltonian for this system is a sum of the
individual grain Hamiltonians and two interaction terms which dynamically
couple the grains: (i) a Coulomb term which accounts for the capacitive
coupling of the electric charges across the TB; and (ii) a tunneling term
responsible for coupling the gap phases of the grains \cite{joe}.
The action for the JJ takes the form
$S_{\scriptscriptstyle JJ}=S_{g}(L)+S_{g}(R)+S_{c}+S_{t}$ \cite{eck}.
Here $S_{g}(i)$ ($i=L,R$) is the single grain action described above
eqn.~(\ref{action}); and
\begin{displaymath}
S_{c}=-\int\,\frac{dt}{\hbar}\frac{C}{2}\left[\,\frac{\hbar}{2e}
       \partial_{t}\gamma\,\right]^{2} \hspace{0.1in} ; \hspace{0.2in}
S_{t}=\int\,\frac{dt}{\hbar}\,\left[\,-I_{c}\frac{\hbar}{2e}\cos\gamma
        \,\right] \hspace{0.1in} .
\end{displaymath}
$C$ is the junction capacitance; $\gamma=(\phi_{L}-\phi_{R})-\int_{1}^{2}
\,{\bf A}\cdot d{\bf l}$ is the gauge invariant (gap) phase difference across
the TB; $I_{c}=\pi\Delta_{0}/2eR_{n}$ is the critical current, and $R_{n}$
is the tunnel junction normal resistance.  We restrict
ourselves to Large JJ's so that localized regions of magnetic flux are
possible (vortices) in the TB, and
the vortex motion in a JJ is one dimensional. The grains are assumed to be
2D (unit thickness in the z-direction), so that the JJ is embedded in
$R^{2}$. Consequently, the TB maps onto a (finite) 1D segment $P\subset
R^{2}$ which will be referred to as the parameter space since
${\bf r}_{0}(t)\in P$ is the set of parameters
appearing in the grain Hamiltonians, and whose time dependence generates
the Berry phases. Since the presence of the WZ term
in the grain action occurs only when the Berry phases are non-vanishing, we
must examine whether this situation is possible in a JJ.

It is enough to consider the Berry phases in the single particle states
since the many-body states will be constructed from them. We first consider
a Traditional (Large Straight) JJ (TJJ) in which the parameter space
$P_{\scriptscriptstyle T}$ is a
1D segment whose endpoints correspond to physically distinct points.
$P_{\scriptscriptstyle T}$ has the topology of the unit interval $I=[0,1]$.
Thus, the only
closed loops $C$ possible in $P_{\scriptscriptstyle T}$ will originate
from
an arbitrary point
$p_{0}$; go out to a point $p_{1}$ via a segment $C_{1}$; and return to
$p_{0}$ along $-C_{1}$. Note that each point on $C_{1}$
is passed through twice. This forces the Berry phase $f_{B}$ to
vanish,
\begin{displaymath}
f_{B}=-\oint_{C_{1}-C_{1}}\,d\tau\,\left[\, i\dot{{\bf r}}_{0}\cdot\langle\,
       E({\bf r}_{0})|\nabla_{{\bf r}_{0}}|E({\bf r}_{0})\,\rangle +
        \frac{e}{\hbar}A_{0}\, \right] = 0 \hspace{0.1in} .
\end{displaymath}
This follows since $\langle\, E|\nabla_{{\bf r}_{0}}|E\,\rangle$
and $A_{0}$ are single-valued functions over $P_{\scriptscriptstyle T}$
so that $\int_{C_{1}}$
is equal and opposite to $\int_{-C_{1}}$ for all $C_{1}$ and for all single
particle states. Consequently, $f_{B}$ vanishes for all the states in the
many-body
Fock space, and no WZ term is generated in $S_{g}(i)$ by the
vortex motion (for a TJJ). From the point of view of Berry's phase, a more
interesting JJ is the Annular JJ (AJJ) in which the TB has a ring-like
topology \cite{AnJJ}. Here the TB maps onto the unit circle in $R^{2}$ so
that $P_{\scriptscriptstyle A}=S^{1}$. Berry phases for closed loops that
wind around $S^{1}$
are non-vanishing (in the gauge $A_{0}=0$, the BCS groundstate Berry phase
induced by a single traversal of $S^{1}$ is $-\pi N_{s}$, where $N_{s}$ is
the number of superconducting electrons).
Thus, for an AJJ, the Berry phases are non-trivial, and the grain actions
each contain a WZ term whose consequences we now explore.

We restrict
ourselves to the case where an applied (uniform) transport current flows
through the AJJ so that ${\bf v}_{s}={\bf v}_{\scriptscriptstyle T}+
{\bf v}_{circ}$. Here ${\bf v}_{\scriptscriptstyle T}$ is the superflow
velocity of the transport current; and
${\bf v}_{circ}$ is that of the superflow circulating about the vortex.
We now show that Berry's phase modifies the current drive acting on a
vortex. This drive
describes the coupling of the grain condensates to the vortex, and the
coupling arises from the terms in $S_{g}(i)$ linear in ${\bf v}_{circ}$.
One such term comes from the $m\rho_{s}{\bf v}_{s}^{2}/2$ term
common to the grain action of all types of JJ's. It leads to the familiar
Lorentz current
drive whose contribution to $S_{t}$ is $\int\, dt(-I\gamma/2e)$
\cite{eck}. In an AJJ, the WZ term also contributes to the current drive
term in
$S_{t}$. We will refer to this contribution as the Magnus drive since its
origin (the WZ term) is the
same as that of the Magnus force \cite{ath,vor}. Let
$\alpha\in S^{1}$ parameterize position along the TB (recall
$P_{\scriptscriptstyle A}=S^{1}$); and
let $\hat{{\bf n}}_{\scriptscriptstyle L}(\alpha )$
($\hat{{\bf n}}_{\scriptscriptstyle R}(\alpha )$) be the unit
outward normal to the face of the Left (Right) grain which interfaces
with the TB. In all case we are familiar with,
$\hat{{\bf n}}_{\scriptscriptstyle L}(\alpha )
=-\hat{{\bf n}}_{\scriptscriptstyle R}(\alpha )=\hat{{\bf n}}(\alpha )$.
By partial integration of these coupling terms, one can show \cite{us}
that the total current drive contribution to
$S_{t}$ is
\begin{equation}
S_{dr}=\int\,dt\, dz\, r d\alpha\,\gamma(\alpha )\,\left[\,
        -\frac{\rho_{s}\hbar}{2} \hat{{\bf n}}(\alpha )\cdot
         ({\bf v}_{T}- \label{drive}
          \dot{{\bf r}}_{0})\, \right] \hspace{0.1in} .
\end{equation}
The ${\bf v}_{\scriptscriptstyle T}$-term is the Lorentz drive, and the
$\dot{{\bf r}}_{0}$-term is the Magnus drive. In principle, the Magnus
drive is non-zero; though, in practice, tunnel junctions are usually
constructed
such that $\hat{{\bf n}}(\alpha )\perp \dot{{\bf r}}_{0}$, so that the Magnus
drive effectively vanishes in a smooth annular geometry. Thus Berry phase
effects in large JJ's are rather subtle. These effects are seen to be
extremely sensitive to the topology of the tunneling barrier, and even where
they are expected to occur (AJJ), the usual tunneling barrier geometry
conspires to nullify its effects!

\underline{{\em (iii) Junction Arrays}:} Finally, we consider a JJA which
corresponds to a lattice $L$
(lattice constant $a_{0}$) whose sites are occupied by SC
grains. Phase coherence of the gap function throughout the array is
established via the Josephson effect. Let $i$ index the lattice sites,
and $\{ n_{i} \}$ index the nearest neighbors of $i$. Vortices in a
JJA live on the dual lattice ${\cal L}$ which is the multiply-connected
region separating the grains, and which is constituted by the TB's
$(i,n_{i})$ between
neighboring grains. Accordingly, the dual lattice is identified with
the parameter space, $P={\cal L}$. With respect to grain $i$, any closed
loop
$C_{i}$ in ${\cal L}$ which winds around $i$ will produce a non-vanishing
Berry phase in this grain's SC groundstate. This will be true for all
grains $i$, so that a WZ term appears in $S_{g}(i)$ for all $i$. The
effective action $S_{\scriptscriptstyle JJA}$ for the array is found by
adding together the action for each tunnel junction
$S_{\scriptscriptstyle JJ}(i,n_{i})$;
\begin{equation}
S_{\scriptscriptstyle JJA}=
\sum_{i}\left\{ S_{0}(i)+S_{wz}(i)+S_{2}(i)+\sum_{\{ n_{i}\} }
  \left[ S_{c}(i,n_{i})+ S_{t}(i,n_{i}) \right] \right\}. \label{array}
\end{equation}
The current drive contibution to $S_{t}(i,n_{i})$ is obtained from
eqn.~(\ref{drive}) by the substitution $\gamma\rightarrow\gamma (i,n_{i})$;
$\hat{{\bf n}}\rightarrow\hat{{\bf n}}(i,n_{i})$. Note that
$\dot{{\bf r}}_{0}$ will not be perpendicular to {\em all\/}
$\hat{{\bf n}}(i,n_{i})$ so that the Magnus drive {\em is\/} active in
a JJA. We now consider the Continuum Limit (CL) for the SC dynamics of the
JJA. Let $l$ be the length scale over which the gauge invariant phase
difference $\gamma(i,n_{i})$ varies; the CL of the SC dynamics corresponds
to $l\gg a_{0}$. In this limit $\gamma(i,n_{i})$ varies little from grain to
grain.  For an array in the CL,
it is reasonable to coarse grain $S_{\scriptscriptstyle JJA}$ over a
length scale $d$ satisfying $a_{0}\ll d\ll l$.
Let the JJA map onto the region $A\subset R^{2}$, and partition
$A$ into cells $dA({\bf x})$ of area $d^{2}$.
In the CL, $\gamma (i,n_{i})$ is essentially constant over each cell.
All grains and junctions lying within $dA({\bf x})$ will be grouped
together in eqn.~(\ref{array}) to give the contribution from this cell
to the coarse grained JJA action $\overline{S}_{\scriptscriptstyle JJA}$.
Because of the JG-eqn, $S_{c}(i,n_{i})$ is seen to represent the action
associated
with the electric field ${\bf E}(i,n_{i})$ present in this junction. A
similar term is present in $S_{2}(i)$ (see eqn.~(\ref{action})) so that
the sum of these two types of terms, for the cell at ${\bf x}$, gives the
action associated with the electric field present in this cell. The coarse
graining of these terms gives the action $dA({\bf x})
[-\overline{{\bf E}}^{2}/8\pi]$ associated with the coarse grained
electric field $\overline{{\bf E}}({\bf x})$ in this cell. One can show
\cite{us} that, in the CL, the integrand of $S_{t}(i,n_{i})$ is
proportional to ${\bf v}_{s}^{2}(i,n_{i})$. A similar term is present in
$S_{2}(i)$ so that the sum of these two types of terms for the cell at
${\bf x}$ gives the action associated with the superflow kinetic energy
in this cell. As with the electric field, coarse graining leads to the
action $dA({\bf x})[m\overline{\rho}_{s}({\bf x})
\overline{{\bf v}}_{s}^{2}({\bf x})/2]$ associated with the coarse grained
superflow $\overline{{\bf v}}_{s}({\bf x})$. Coarse graining $S_{0}(i)$,
$S_{wz}(i)$, and the remaining terms in $S_{2}(i)$ is straightforward as
they are essentially constant over a cell $dA({\bf x})$ and so yield terms
of similar form, but with coarse grained quantities appearing in the final
result. We find,
\begin{displaymath}
S_{\scriptscriptstyle JJA}\rightarrow \overline{S}_{\scriptscriptstyle JJA}
 =S_{0}+\int_{A}\,d^{2}x dt\,\left[\, \overline{\rho}_{s}\left(
   \frac{\hbar}{2}
  \partial_{t}\phi +e\overline{A}_{0}\right)+N(0)\left(\frac{\hbar}{2}
   \partial_{t}\phi +e\overline{A}_{0}\right)^{2}+
    \frac{m\overline{\rho}_{s}}{2}\overline{{\bf v}}_{s}^{2}+\frac{1}{8\pi}
     \left\{ \left( \overline{{\bf H}}-{\bf H}_{ext}\right)^{2}-
      \overline{{\bf E}}^{2}\right\} \, \right] .
\end{displaymath}
Thus $\overline{S}_{\scriptscriptstyle JJA}$ has exactly the same form as
the action of a single bulk 2D superconductor $S_{g}$
(see eqn.~(\ref{action})), so that, in the CL,
the SC dynamics of a JJA maps onto that of a 2D SC film.
Because the Berry phases are non-trivial
in the array, $\overline{S}_{\scriptscriptstyle JJA}$ will contain a
WZ term, and consequently, the non-dissipative force ${\bf F}_{nd}$
acting on a vortex will
contain a Magnus force contribution, exactly as in the 2D
film \cite{ath,vor}.
We stress that the appearance of the Magnus force in the dynamics of
vortices in a JJA follows naturally from the presence of a Berry phase in
the SC dynamics; it is not necessary to assume that external
charges are fixed to the grains to generate the Magnus force as done in
Ref.~\cite{faz}. Having established
that the vortex dynamics of a JJA maps onto that of a 2D SC film, we must
point out an important advantage of arrays in the CL over 2D films for
studies of vortex dynamics. Because the array lattice can be constructed
with very great precision, one can control the degree of disorder present
in a particular array. This raises the possibility of carrying out
flux-flow experiments in arrays (in the CL) in which the degree of disorder
can be systematically varied from (essentially) zero to any specified
degree, or distribution, of disorder. This is usually {\em not\/} the
case in 2D
films. As an example, consider the Hall Effect in an array in the CL with
negligible disorder. The only forces acting on a vortex will be the
non-dissipative force ${\bf F}_{nd}=a{\bf v}_{s}\times\hat{{\bf z}}-
b{\bf v}_{L}\times\hat{{\bf z}}$ ($a$, $b$ being determined
via the coarse grain averaging); and the dissipative force
$-\eta {\bf v}_{L}$. In the steady state, the sum of these forces is zero,
allowing a determination of the Hall angle, $\tan\Theta_{H} =-v_{Lx}/v_{Ly}
=b/\eta$. This is exactly the result one finds in a 2D SC film \cite{vin}.
Thus
one expects a sign anomaly in the Hall Effect to occur in JJA's (in the CL)
when the grains are made of a material which shows a sign anomaly in the
2D film experiments. The ability to systematically vary disorder
in arrays would appear to make them ideally suited for testing two recent
predictions of Vinokur et.\ al.\ \cite{vin}: (i) independence of the
Hall conductivity $\sigma_{xy}$ on disorder; and (ii) power law behavior of
the Hall resistivity $\rho_{xy}\sim\rho_{xx}^{2}$ after averaging over
disorder.

We would like to thank the International Atomic Energy
Agency; UNESCO; and ICTP, Trieste for support. F. G. would like to thank
T. Howell III for constant support.


\begin{references}
\bibitem{ath} P. Ao and D. J. Thouless, Phys.\ Rev.\ Lett.\ {\bf 70},
              2158 (1993).
\bibitem{con} Regrettably, what is meant by the ``Magnus Force'' is not
              uniformly agreed upon in the literature of Type-II
              superconductors. In this paper we will follow the terminology
              introduced in Ref.~\cite{vor}.
\bibitem{go3} M. Stone, {\em Bosonization\/} (World Scientific, New
              Jersey, 1994), chap.\ 3;
              I. J. R. Aitchison, P. Ao, D. J. Thouless, and X.-M. Zhu,
              Phys.\ Rev.\ B {\bf 51}, 6531 (1995).
\bibitem{vor} F. Gaitan, Phys.\ Rev.\ B {\bf 51}, 9061 (1995).
\bibitem{ram} Thermodynamic manifestation of Berry's phase in the
              magnetization of superconducting networks in the mixed state
              has been discussed in R. Rammal, Physica B {\bf 152}, 37
              (1988).
\bibitem{us}  F. Gaitan and S. R. Shenoy, in preparation.
\bibitem{vin} V. M. Vinokur, V. B. Geshkenbein, M. V. Feigel'man, and
              G. Blatter, Phys.\ Rev.\ Lett.\ {\bf 71}, 1242 (1993).
\bibitem{eck} U. Eckern, G. Schon, and V. Ambegaokar, Phys.\ Rev.\ B
              {\bf 30}, 6419 (1984).
\bibitem{and} P. W. Anderson, Rev.\ Mod.\ Phys.\ {\bf 38}, 298 (1966).
\bibitem{joe} B. D. Josephson, Phys.\ Lett.\ {\bf 1}, 251 (1962). Damping
              due to quasiparticle tunneling
              is also introduced by the tunneling interaction; it will
              be ignored as it does not affect our arguments.
\bibitem{AnJJ} B. Dueholm et.\ al.\ , in {\em Proceedings of the Seventeenth
              International Conference on Low Temperature Physics\/ },
              eds.\ U. Eckern, A. Schmid, W. Weber, and H. W\"{u}hl (North
              Holland, Amsterdam 1984), p.\ 691.
\bibitem{faz} R. Fazio et.\ al.\ , Helv.\ Phys.\ Acta {\bf 65}, 228 (1992).
\end{references}
\end{document}